\documentclass[a4paper,graphics,natbib,graphicx,psfig,amssymb,ccaption]{article}
\usepackage{lscape}
\usepackage{graphicx}
\newcommand{\affil}[1]{$^{\rm #1}$}
\usepackage[authoryear]{natbib}
\bibpunct{(}{)}{;}{a}{}{,}
\date{} 

\title{\large\bf\flushleft Australian participation in the Gaia Follow-Up Network for Solar System Objects }
\author{\parbox{\textwidth}{\flushleft
\vspace{-0.5cm}
{\it M.~Todd\affil{A,E}, D.~M. Coward\affil{B}, P. Tanga\affil{C} W. Thuillot\affil{D}}\\
\vspace{0.4cm}
 {\small \affil{A}\,Department of Imaging and Applied Physics, Bldg 301, Curtin University of Technology,\\
Kent St, Bentley, WA 6102, Australia}\\
{\small \affil{B}\,School of Physics, M013, The University of Western Australia, 35 Stirling Hwy,\\
Crawley, WA 6009, Australia}\\
 {\small \affil{C}\,UMR 6202 Cassiop\'ee, University of Nice-Sophia Antipolis, CNRS, Observatoire de la C\^ote
d'Azur, BP. 4229, 06304 Nice Cedex 4, France}\\
 {\small \affil{D}\,IMCCE, Observatoire de Paris, CNRS, UPMC, Univ. Lille 1, 77 avenue Denfert-Rochereau, 75014 Paris, France}\\
{\small \affil{E}\,Corresponding author. E-mail: michael.todd@icrar.org}}} 

\begin{document}
\twocolumn[
\begin{changemargin}{.8cm}{.5cm}
\begin{minipage}{.9\textwidth}
\vspace{-1cm}
\maketitle
%
%
\small{ {\bf Abstract:} The Gaia satellite, planned for launch by the European Space Agency (ESA) in 2013, is the next generation astrometry mission following Hipparcos. Gaia's primary science goal is to determine the kinematics, chemical structure and evolution of the Milky Way Galaxy. In addition to this core science goal, the Gaia space mission is expected to discover thousands of Solar System Objects.  Because of orbital constraints Gaia will only have a limited opportunity for astrometric follow-up of these discoveries. In 2010, the Gaia consortium DPAC initiated a program to identify ground-based optical telescopes for a Gaia follow-up network for Solar System Objects to perform the following critical tasks: confirmation of discovery, identification of body, object tracking to constrain orbits. To date this network comprises 37 observing sites (representing 53 instruments). The Zadko Telescope, located in Western Australia, was highlighted as an important network node because of its southern location, longitude and automated scheduling system. We describe the first follow-up tests using the fast moving Potentially Hazardous Asteroid 2005 YU55 as the target. }


\medskip
\medskip
\end{minipage}
\end{changemargin}
]
\small

%
\section{Introduction}
\label{sectintro}
The European Space Agency (ESA) Gaia mission, scheduled for launch in Spring 2013, is a space-based all-sky survey. The Gaia spacecraft will provide astrometry, photometry and spectroscopy for point-like sources down to $V\sim20$. Gaia's science data comprises absolute astrometry, broad-band photometry, and low-resolution spectro-photometry. During its 5-year mission, Gaia will survey about 1 billion stars and 300,000 Solar System objects, of which the majority will be Main Belt asteroids. It will also survey about 500,000 point-like extragalactic sources and $\sim1$ million faint galaxies. The astrometric precision for the mission will be better than $10\mu as$ for stars brighter than $V\sim13$ and about $25\mu as$ for stars $V\sim15$. Gaia will initially build on, and then surpass, the results of the Hipparcos mission of about 20 years ago \citep{Mignard11a}. Such an observational effort has been compared to mapping the human genome, for the amount of collected data and for the impact that it will have on all branches of astronomy and astrophysics. 

In addition to the above core science goal, the Gaia space mission will discover exotic transient objects in large numbers. Many thousands of transients will be discovered including exoplanets and supernovae. Tens of thousands of brown and white dwarfs will be identified spectroscopically and, within our Solar System, some hundreds of thousands of minor planets will be observed. Of particular interest will be the numbers of unusual minor planets such as minor planets which have high inclinations, so that they are normally outside the regions of sky routinely surveyed by Near-Earth Asteroid (NEA) programmes, and inner Solar System Trojan asteroids.

Because Gaia's primary mission is to perform a space-based all-sky survey, it is not designed to conduct any targeted follow-up studies. Gaia will be constrained by its orbit and by its design to survey the sky as completely as possible \citep{2007EM&P..101...97M}. For these reasons, and because of the expected vast number of discoveries of transient phenomena, Gaia will not be able to either confirm or perform detailed studies of the discoveries. 

\subsection{Gaia Follow-Up Network}

During its 5 year mission, Gaia will observe many new SSOs (Solar System Objects). One feature of Gaia observations
is the ability to image at rather low solar
elongation (45 degrees), enabling detections of Earth-crossing asteroids (Atens) and Inner Earth Asteroids (IEAs), and discoveries of new NEAs at larger solar elongation. In performing an all-sky survey Gaia will necessarily survey regions of sky away from the regions targeted by NEA programmes. This holds the potential to discover asteroids with high orbital inclinations, and Trojan asteroids in the stable Lagrangian regions in the orbits of the planets in the inner Solar System \citep{2011arXiv1111.2427T,2012MNRAS.420L..28T,2012MNRAS.424..372T}.

In such cases, due to the motion of the objects and limiting magnitude, the
scanning law of Gaia will restrict the orbit determination to be constrained on a very small
number of astrometric measurements. This implies that only a ground-based
optical network can obtain accurate orbital modeling, based on enough astrometric
measurements. This is the primary objectives of the Gaia Follow-Up
Network for Solar System Objects (FUN-SSO).

In addition to the improvement in accuracy of the astrometric data used for orbital modeling of 
specific objects, some rare and peculiar SSOs such as asteroids with cometary activity could be studied \citep{2008P&SS...56.1812T}. Due to the limitations of Gaia's observing method, a ground-based follow-up network will be crucial for studying the physical characteristics of these objects.

In order to be effective in acquiring high-quality follow-up imaging following an alert by the Gaia data processing system, the network must have a large geographical coverage. This is why several observing stations have been invited for participation in this project. In 2010, the Gaia DPAC (Data Processing and Analysis Consortium) initiated a program to indentify optical telescopes for a Gaia follow-up network for SSOs to perform the following critical tasks: confirmation of alert, identification of transient, continuous monitoring and tracking. At the Gaia Follow-up Network Meeting in Paris 2010, the Zadko Telescope was highlighted as an important contribution to this network, because of its longitude and robotic operation. To date the network comprises 37 observing sites (representing 53 instruments) (cf. figure 1). 

 \begin{figure*}
 \begin{center}
 \includegraphics[scale=0.80]{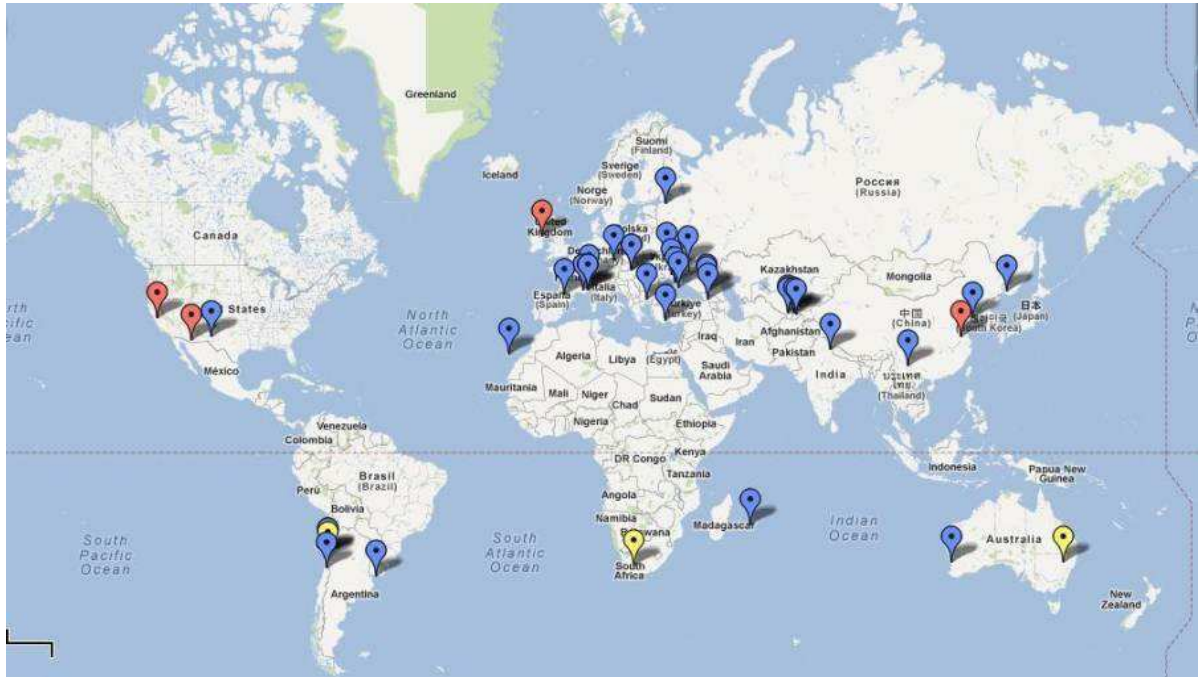}
 \caption{ \small
  The global distribution of observing stations participating in the Gaia Follow-Up Network of Solar System Objects. The distribution is biased towards northern latitudes and European longitudes. Blue markers are the current observing stations, red markers are existing facilities which have expressed interest but are not confirmed, and yellow markers are planned future facilities.}
 \end{center}
 \end{figure*}

SSOs and stars are not differentiated on board but are treated the same way. Each source brighter than $V = 20$ is identified by the star mapper (SM) on the CCD array. A window is defined for each identified source for accumulation of charge as it crosses the CCD. An automated data reduction pipeline (implemented by the science consortium, DPAC) is run over all sources. Those which are found to move (from one epoch to the other, essentially) or are extended are sent to a specific branch of the pipeline which applies a specialized data reduction including identification and processing of SSOs. Most of the SSOs that Gaia will observe will be known, so they will be simply identified by comparison to ephemeris derived from the complete Minor Planet Center (MPC) catalogue. The remaining objects which are moving and not identified are classed as new asteroid candidates. For these, a tentative orbit is computed for ground-based recovery and an alert will be issued to the Gaia-FUN-SSO network.

Alerts from the Gaia data processing system will be received between 24 and 48 hours after detection.  Problems may arise for observations of peculiar objects: fast moving objects, faint objects, NEAs close to Earth with strong parallaxes. Hence, small and less sensitive telescopes ($<0.6$ m) can be
useful but more sensitive instruments ($>1$ m) will be essential. Furthermore, fully robotic telescopes are proven as ideal for observations triggered by an alert. Among the telescopes in the Gaia-FUN-SSO network, five are robotic telescopes\footnote{\url{http://www.imcce.fr/gaia-fun-sso/} for an updated list}.

 \section{The Zadko Telescope}
\label{sectzadkotel}
The Zadko Telescope\footnote{\url{http://www.zt.science.uwa.edu.au}}
is a 1-m f/4 Cassegrain telescope,
built by DFM Engineering,
Inc.\footnote{\url{http://www.dfmengineering.com}},
and is situated in the state of Western Australia
at longitude $115^{\circ} 42' 49''$ E,
latitude $31^{\circ} 21' 24''$ S,
and at an altitude of $47$\,m above sea level ($67$\,km north of Perth).

It is an equatorial, fork-mounted telescope
with a primary mirror clear aperture of $1007.0$\,mm,
and a system focal length of $4038.6$\,mm.
Its fast optics have a low f-ratio, coupled with
a flat, wide field of view (maximum possible FoV$ \approx 1^{\rm o}$).

The core science theme of the Zadko Telescope is the discovery and study of transients \citep{2010PASA...27..331C}. The facility was robotized in 2010 employing an observatory control system developed for the TAROT telescopes. It can acquire photometric data autonomously on gamma ray bursts (GRBs) and Galactic transients following alerts by the Swift satellite and other external triggers. In addition to the follow-up of GRB triggers, the Zadko Telescope is used for follow-up of exo-planet discoveries, follow-up imaging of ANTARES alerts, follow-up imaging of LIGO/Virgo gravitational wave alerts and SSOs. Below is a list of some Zadko Telescope observations.
\begin{itemize}

\item Zadko measures light curve of GRB101024 only 200 seconds after satellite trigger, revealing unexplained optical light curve features (Gendre et al. 2011).
\item In 2009, a minor planet pilot search program resulted in the discovery of 13 uncatalogued Solar System objects (SSO). Further studies resulted in four of these being linked to `orphan' observations in the MPC database.
\item 14 follow-ups of Swift satellite GRB triggers with NASA Gamma-ray Coordinates Network notices.
\item TAROT + Zadko network successfully image sky positions triggered by LIGO for the detection of possible optical counterparts to gravitational wave candidates \citep{2012A&A...539A.124A}.
\item GRBs 090205 and 090516, with respective redshifts of 4.3 and 4.1, are among the most distant optical transients imaged by an Australian telescope.
\end{itemize}

For Gaia to achieve astrometric accuracy on scales of a few tens of micro-arcseconds its reference frame must be well-defined. By using the optical positions of QSOs, assuming that QSOs have no peculiar transverse motion, then any observed proper motion will reveal the global rotation of the Gaia sphere. Along with the robotic TAROT telescopes in France and Chile, the Zadko Telescope has been observing selected QSOs since mid-2011 to assist in constructing the initial catalogue for establishing Gaia's reference frame. 

The Zadko Telescope observatory is being upgraded in 2012. The observatory building is being replaced with a new observatory optimized for automated rapid response imaging of transients and SSOs. Furthermore a new instrument and imaging package with spectroscopic capability, funded by the Australian Research Council, will be installed in 2012-2013.

 The capability of the Zadko telescope to detect faint objects to $m\approx21.5$, and its robotic scheduling system enables it to perform optimal targeted follow-up observations. Since 2010, routine examination of selected image sets where the cadence has facilitated detection of moving objects has resulted in the detection of 11 additional uncatalogued SSOs. Follow-up studies have enabled three of these to be linked to earlier detections in the MPC database. In some cases these linkages have been made between apparitions at several previous oppositions over a period of years up to a decade (see Table \ref{tabasteroids}). 

\begin{table}[tbh!]
\begin{center}
\caption{ \small Zadko Telescope asteroid discovery and follow-up statistics.}
\label{tabasteroids}
\begin{tabular}{llll}
\hline
Year        & Detections & No. of objects  & Link span  \\
            &            & linkages & (opp.s/years)   \\
\hline
2009           & 13      & 4           &  4 / 10  \\
2010           & 4       & 1           &  3 / 5   \\
2011           & 6       & 1           &  5 / 5   \\
2012           & 1       & 1           &  2 / 6   \\
\hline
\end{tabular}
\end{center}
\end{table}

\section{Preliminary Gaia FUN-SSO testing}

In November 2011, observations were made of the Potentially Hazardous Asteroid 2005 YU55. On November 8 this object passed within 325~000~km of Earth\footnote{\url{www.jpl.nasa.gov/news/news.cfm?release=2011-332}}. Observing this target so soon after its fly-by of Earth meant that its position was rapidly changing (see Figure 2). Observation of this target demonstrated the capability of the Zadko Telescope to accurately point and acquire images of a relatively fast-moving object. On the first night of observation the sky motion was $\sim 4$ arcseconds/minute. By comparison, main-belt asteroid sky motions are typically smaller than $0.5$ arcseconds/minute.

The Zadko Telescope made observations of 2005 YU55 over a period spanning 11 nights, from November 11 to 21. This was the third-longest observational coverage of 14 participating telescopes, and was the only contribution from a Southern Hemisphere telescope. The Zadko Telescope was used to obtain 42 images of PHA 2005 YU55 in November 2011, of which 21 were used for astrometry. The images and measurements were delivered to the Gaia FUN-SSO working group, which reported the collective results from participating instruments to the MPC. During this time the target faded dramatically (from 13.7 to 18.1 in SDSS r filter).

The residual error between reported (observed) position and the position calculated by the Gaia FUN-SSO working group was calculated by taking $\sim 3600$ observations between December 2005 and December 2011 and fitting a numerical model of the asteroid motion. The model takes into account perturbations from all eight major planets and Pluto and the relativistic effect of the Sun. The mean residual was determined to be 0.30 arcseconds, and for the Zadko Telescope was typically better than 0.20 arcseconds in both Right Ascension and Declination. The first measurement from each night, with residuals (O-C values), is given in Table \ref{tabk05y55u}, and shows the change in brightness and sky motion over that period.

\begin{table*}[tbh!]
\begin{center}
\caption{ \small Selection of 2005 YU55 observations made by the Zadko Telescope showing the change in brightness and sky motion over the period. The O-C values are the differences from the calculated positions.}
\label{tabk05y55u}
\begin{tabular}{llllllr@{.}lr@{.}l}
\hline
Date (UT)  & R.A.  & Decl.  & Mag. & \multicolumn{2}{c} {Sky Motion} & \multicolumn{4}{c}{O-C}  \\
           &       &        &      & \multicolumn{2}{c}{(arcsec/min)} & \multicolumn{4}{c}{(arcsec.)}  \\
           &       &        &      & \multicolumn{1}{c}{$\alpha$}  & \multicolumn{1}{c}{$\delta$}  & \multicolumn{2}{c}{$\alpha$.cos($\delta$)}   & \multicolumn{2}{c}{$\delta$}   \\
\hline
2011 11 11.60593 & 02 13 56.45 & +17 24 22.7  & 13.7 & +4.1 & -0.9 & -0&43  & -0&10  \\
2011 11 12.60797 & 02 20 47.59 & +17 06 54.4  & 14.2 & +1.9 & -0.5 & -0&20  & -0&09  \\
2011 11 15.60574 & 02 29 03.05 & +16 45 08.8  & 16.1 & +0.3 & -0.1 &  0&05  & -0&12  \\
2011 11 16.60259 & 02 30 24.47 & +16 41 42.2  & 16.3 & +0.2 & -0.09 &  0&01  &  0&04  \\
2011 11 19.59546 & 02 33 06.84 & +16 35 40.7  & 17.7 & +0.03 & -0.03 & -0&00  & -0&13  \\
2011 11 21.59059 & 02 34 22.04 & +16 33 47.7  & 18.3 & +0.02 & -0.02 &  0&50  & -0&02  \\

\hline
\end{tabular}
\end{center}
\end{table*}

\begin{figure}[tbh!]
\includegraphics[scale=0.95]{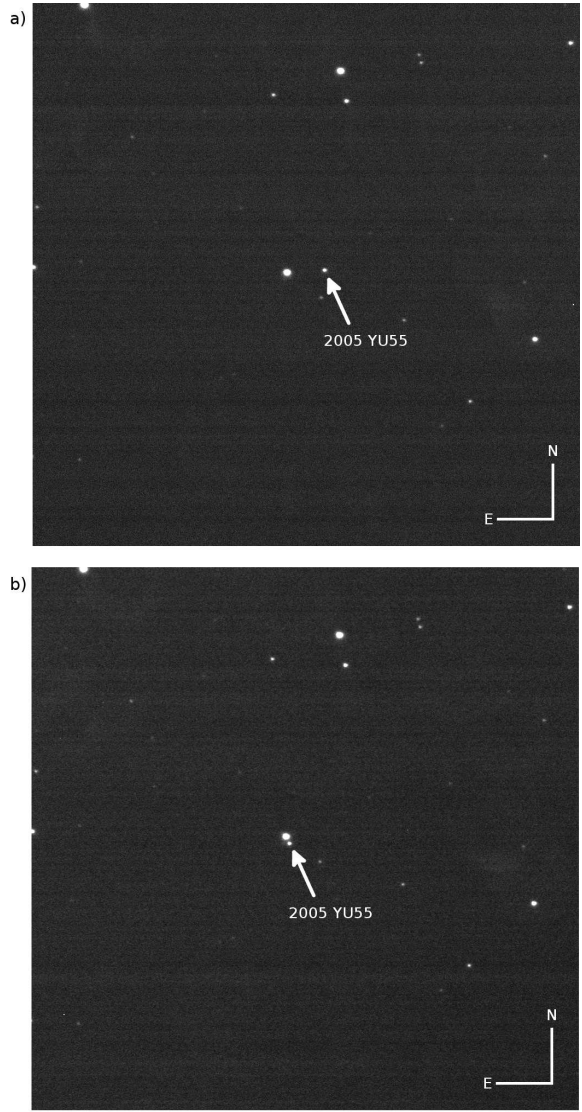}
\caption{ Zadko Telescope images of Potentially Hazardous Asteroid 2005 YU55. Exp: 60s, Field centre: 02h20m49s6 +17d06m47s, Scale: 11.5 x 11.5 arc-minutes, Date: 2011-11-12.60964 (a) and 2011-11-12.62607 (b).  }
\end{figure}

In February 2012, observations were planned for asteroids 99942 Apophis and 1996 FG3. 
Apophis was selected since it will have a very close approach in 2029 and new astrometry would be useful to obtain more accurate ephemerides. New data can also be used improve our knowledge of the Yarkovsky effect on its trajectory.
1996 FG3 was selected as it is the main target of the proposed Marco Polo-R space mission which will provide a sample return. Better knowledge of its orbital characteristics will help with the success of the mission.
Although no data were obtained for these because of observing constraints, the targets were successfully logged into the robotic scheduler as a test.

\section{Summary}
During its 5 year mission, Gaia will discover new SSOs, including inner earth asteroids (IEAs) and Earth-crossers (Atens), and new NEAs at larger solar elongations. Because the scanning law of Gaia will restrict the orbit determination to a very small
number of astrometric measurements, only a ground-based
optical network can obtain accurate orbital modeling, based on enough astrometric
measurements. This is the primary objective of the Gaia Follow-Up
Network for Solar System Objects (FUN-SSO).

In 2010, the Gaia DPAC (Data Processing and Analysis Consortium) initiated a program to identify ground-based optical telescopes for the Gaia FUN-SSO. The main criteria for the network are facilities that are ideally robotic, or can be re-scheduled to give priority to a Gaia alert. The Zadko Telescope was identified as a potentially important contributor to the network because of its robotic operation, sensitivity and location. It will used for validation of new SSOs, astrometry and photometry of these targets.

In 2011, the Gaia FUN-SSO working group initiated an observing campaign of SSOs, with the main goal to test the capability of the network node facilities. Simulated Gaia alerts were distributed to the network (initially by email) providing ephemerides for follow-up imaging of real targets.
We tested the Zadko Telescope capability for participating in the observing campaign of PHA 2005 YU55 in late 2011. Of the 37 observing stations in the network, 11 were able to participate and provide useful astrometric data. The Zadko Telescope contributed the third-longest observational coverage from 14 participating telescopes, and was the only contribution from a Southern Hemisphere telescope. The data from the observations showed that the Zadko Telescope scheduling system worked successfully, with accurate telescope pointings, and positional data that was used for astrometry with an uncertainty of about 0.2 arcseconds. Testing of the networks ability to constrain orbits of SSOs will continue until the launch of Gaia.

Following upgrades to the Zadko Telescope observatory in 2012-2013, the telescope will resume operation several optical transient projects. These improvements will also provide scope for participation in additional projects, including follow-up of exo-planet candidates. Furthermore, an image processing pipeline for automated identification and analysis of optical transients is in development and planned for implementation in 2013.

%
%
%
%
\section*{Acknowledgements} 
D.~M. Coward is supported by an Australian Research Council Future Fellowship. The Zadko Telescope was made possible by a philanthropic donation to the University of Western Australia by the Zadko family.
%


\end{document}